\documentstyle[prl,aps,preprint]{revtex}
\begin{document}
\draft
\title{Field-induced magnetic anisotropy in La$_{0.7}$Sr$_{0.3}$CoO$_3$}   
\author{J. Mira and J. Rivas}
\address{Departamento de F\'\i{}sica Aplicada, Universidade de
Santiago de Compostela, E-15782 Santiago de Compostela, Spain}
\author{M. V\'azquez}
\address{Instituto de Ciencia de
Materiales, CSIC. E-28049, Madrid, Spain}  
\author{M. R. Ibarra}
\address{Departamento de F\'\i{}sica de la Materia Condensada-ICMA, Univesidad de
Zaragoza-CSIC, E-50009 Zaragoza, Spain} 
\author{R. Caciuffo}
\address{Istituto Nazionale per la Fisica della Materia and Dipartimento di
Fisica ed Ingegneria dei Materiali, Universit\'a di Ancona, I-60131 Ancona,
Italy}  \author{M. A. Se\~nar\'\i{}s Rodr\'\i{}guez} 
\address{Departamento de Qu\'\i{}mica
Fundamental e Industrial, Universidade da Coru\~na,  E-15071 A Coru\~na, Spain}

\maketitle

\begin{abstract}

Magnetic anisotropy has been measured for the ferromagnetic 
La$_{0.7}$Sr$_{0.3}$CoO$_3$ perovskite from an analysis of the high-field part 
of the magnetization vs. field curves, i.e., the magnetic saturation regime.
These measurements give a magnetic anistropy one order of magnitude higher
than that of reference manganites. Surprisingly, the values of the
magnetic anisotropy calculated in this way do not coincide with those estimated
from measurements of coercive fields which are one order of magnitude
smaller. It is proposed that the reason of this anomalous behaviour is a
transition of the trivalent Co ions under the external magnetic field from a
low-spin to an intermediate-spin state. Such a transition converts the Co$^{3+}$
 ions into Jahn-Teller ions having an only partially quenched orbital
angular momentum, which enhances the intra-atomic spin-orbit coupling and
magnetic anisotropy.

\end{abstract}


Attention to cobalt perovskites of formula {\it R}$_{1-x}${\it A}$_x$CoO$_3$
was first paid in the 50's. At that time, Jonker and van
Santen \cite{Jonker} and Koehler and Wollan \cite{Koehler}
described their basic properties, that were first interpreted by
Goodenough \cite{Goodenough}.  The finding of colossal magnetoresistance effects
in Mn perovskites \cite{VH,Jin} reactivated the interest on these and
related materials, like the title compound. Since the first
investigations it was seen that the magnetic and transport properties of cobalt
perovskites have a peculiar
thermal dependence \cite{Koehler,Heikes,Raccah,Menyuk,Bhide,Tona1}. To explain
it, the existence of a thermally induced change in the electronic configuration
of the Co ions was established by Raccah and Goodenough \cite{Raccah}. It is a
consequence of the interplay between the intra-atomic exchange and the crystal
electric field interactions, which in this particular compound are of
comparable magnitude. This fact is highlighted as the source of many of the
differences with manganites \cite{MiraPRB}, where the intra-atomic exchange is
rather higher than the crystal electric field interaction. For the particular
case of LaCoO$_3$, the electronic state changes at intermediate temperatures. At
low temperatures, the predominant state is low-spin ($t_{2g}^6$, S=0) and, at
intermediate ones, it evolves to an intermediate- ($t_{2g}^5$$e_g^1$, S=1) or
high-spin state 
($t_{2g}^4$$e_g^2$,
S=2)
\cite{Raccah,Menyuk,Tona1,Asai,Korotin,Saitoh,Yamaguchi,Zhuang,Tokura,Yamaguchi2},
due to the small energy gap ($\simeq 0.03$ eV) between such configurations.
Sr-doping leads to a phase segregation into hole-rich ferromagnetic clusters and
a hole-poor LaCoO$_3$-like matrix \cite{Tona2,CaciuffoPRB}. The hole-rich regions
are isolated for low x, but at $x \simeq$ 0.20 they percolate and the samples
become ferromagnetic and metallic. The phase diagram of this very complex system
has been given by  Se\~nar\'\i{}s Rodr\'\i{}guez and Goodenough \cite{Tona2}.

One of the specific features of La$_{0.7}$Sr$_{0.3}$CoO$_3$ is the difficulty of
approaching a magnetic saturation state \cite{MiraPRB,Tona2,Itoh,Vazquez}. This
has been attributed to the absence of true long-range-order, that leads to a
cluster-glass state \cite{Itoh}. Nevertheless, the ultimate origin of the
situation is still not clear, and some hypotheses have been launched
recently, like the presence of diamagnetic S=0 Co ions that
would dilute the magnetic lattice \cite{MiraPRB}. In any case, after the
percolation of the hole-rich ferromagnetic clusters the system can be
considered as a ferromagnet.

As a new input to this puzzle, Ibarra {\it et al.} reported recently a huge
anisotropic magnetostriction \cite{Ibarra} that cannot be explained on the basis
of the usual spin-orbit coupling contribution. To explain it, they proposed an
orbital instability of Co$^{3+}$ under an external magnetic field. In this
work the hypothesis of the orbital instability is adopted to analyze the high
magnetic anisotropy of the La$_{0.7}$Sr$_{0.3}$CoO$_3$ system.

The sample was prepared by conventional ceramic methods. X-ray powder
diffraction revealed that it was single phase. Initial magnetization data were
taken with vibrating-sample and SQUID magnetometers in fields up to 55 kOe.
Before each run, the sample was demagnetized heating up to 350 K, well above
the Curie point, and cooled down to the test temperature at zero field.

In Fig. \ref{un} the high-field part of the magnetization vs. magnetic field
isotherms is shown. As can be seen, the difficulty to achieve saturation is
greater than in ferromagnetic manganites. In
order to analyze the approach to saturation, magnetization we use the empirical
relationship \cite{Bozorth,Morrish}

\begin{equation}
M_H = M_S(T) - \frac{a}{H} - \frac{b}{H^2} + \chi H
\label{ec1}
\end{equation}
where M$_H$ is the component of the magnetization along the field direction,
M$_S$ is the saturation magnetization and {\it a}, {\it b} and $\chi$ are
constants. The term $\chi H$ represents the field-induced increase in the
spontaneous magnetization of the domains and it is very small at temperatures
well below the Curie temperature. The $a/H$ term is generally interpreted as
due to imperfections such as dislocations or nonmagnetic inclusions, and
$b/H^2$ is due to crystalline anisotropy. To expedite the fits, the derivative is
often used \cite{Bozorth}

\begin{equation}
(\frac{dM_H}{dH} - \chi) H^3 = aH + 2b.
\label{ec2}
\end{equation}

Data were fitted to achieve the $\chi$ that best linearizes the curves. In Fig.
\ref{dous}(a) we show the representation of the magnetization according to Eq.
\ref{ec2}, with the $\chi$'s shown in Fig. \ref{dous}(b). Linear fits were then
done for data above 30 kOe in order to obtain $a$ and $b$. These data,
substituted in Eq. \ref{ec1}, allow to fit the magnetization data and 
to determine the saturation magnetization shown in Fig. \ref{dous}(c).

From the value of $b$, the values of
the anisotropy constant can be calculated. For a polycrystalline
material with crystallites oriented at random, the approach to saturation may be
calculated by averaging the magnetization curves for all crystal
orientations \cite{Herpin}. The {\it b} constant is then given by
\cite{Holstein}

\begin{equation}
b = \beta \frac{K^2}{M_S}
\label{ec4}
\end{equation}
where $\beta$ comes from purely geometric
considerations. As this study only estimates the
orders of magnitude of the magnetic anisotropy, we chose the same value of
$\beta=8/105$ used for the reference manganese perovskite
La$_{2/3}$Sr$_{1/3}$MnO$_3$ (Ref. \cite{Balcells}), understanding that the
choice does not affect the final results. The values thus obtained for the
anisotropy constant are reported in Fig. \ref{tres}. 

The effects of magnetocrystalline anisotropy are detectable in other features of
ferromagnetic materials, like the value of coercive fields. The coercive field
can be estimated from the results of Fig. \ref{dous} with the well-known
relation

\begin{equation}
H_C(T) = A \frac{K}{M_0(T)}
\label{ec5}
\end{equation}
where A is a constant tipically between 0.1 and 1 (Ref. \cite{Chikazumi}) and
M$_0$ is the spontaneous magnetization. Taking
into account that for T $<$ 0.8T$_C$ M$_0 \simeq$M$_S$ (M$_S$ can be used
instead of M$_0$ at low temperatures), we arrive to the fact that expected
coercive fields are of the order of 10$^3$ or 10$^4$ Oe. The coercive fields of 
La$_{0.7}$Sr$_{0.3}$CoO$_3$ have already been reported by V\'azquez {\it et
al.} \cite{Vazquez}; surprisingly, a clear discrepancy is found with the
values expected from Eq. \ref{ec5}, in one or two orders of magnitude, depending
on the A value.

A deeper insight is needed, and the value of A is important to quantify this
anomaly. In order to determine A, we have analyzed the manganite
La$_{0.7}$Ca$_{0.3}$MnO$_3$. From magnetization vs. field isotherms at 5 K and
fitting the high-field part in the same way as previously, we obtain K= $3.4
\cdot 10^5$ $erg/cm^3$ ($5.86 \cdot 10^4$ $erg/g$), which is consistent
with the results of Perekalina {\it et al.} \cite{Perekalina}, who give values
ranging from  $3.5 \cdot 10^4$ to $6 \cdot 10^4$ erg/g for the magnetic
anisotropy of La$_{0.7}$(Sr,Pb)$_{0.3}$MnO$_3$. It is worth mentioning that this
value is one order of  magnitude  smaller than that of the cobaltite.
Provided that coercive fields at 5 K are of the order of
$\sim$ 50 Oe and M$_S \sim 650$ $emu/cm^3$, we arrive to the conclusion that
A$\simeq$ 0.1, which will be used for the analysis of
La$_{0.7}$Sr$_{0.3}$CoO$_3$. With this value of A and the data of H$_C$
and M$_S$, we can recalculate the anisotropy constant by Eq. \ref{ec5}. The
result is shown in Fig. \ref{tres}. A simple comparison of these data (measured
from coercive fields, i.e., at low applied magnetic fields) with Fig. \ref{tres}
(obtained from the high-field regime) show that they are around one order of
magnitude smaller. 

The question now is to identify what is happening with the magnetic anisotropy
of this compound. For the measurements of the two sets of anisotropy constant
values (expected in principle to be similar), the only external parameter that
has been changed is the external magnetic field. In one case it is high
(saturation regime) and in the other it is low (when measuring coercive fields,
that are of the order of $10^2$ Oe). Then, it seems that the magnetic
anisotropy constant of La$_{0.7}$Sr$_{0.3}$CoO$_3$ does not remain the same under
external magnetic fields. We see a strong parallelism of this situation with
the finding by Ibarra {\it et al.} \cite{Ibarra} of a huge anisotropic
magnetostriction induced by the external magnetic field. They concluded that
the anisotropic magnetostriction cannot be explained on the basis of the usual
spin-orbit contribution and proposed that it is due to the orbital
instability of the Co ions under the external applied field; i.e., on the
transformation of the low-spin state of the Co$^{3+}$ ions in the
insulating matrix (which is diamagnetic and does not contribute to magnetic
anisotropy) into the energetically close intermediate-spin configuration, which
is a Jahn-Teller ion. As described by Ibarra {\it et al.} \cite{Ibarra}, under
Jahn-Teller distortion, the doubly degenerated $e_g$ level of IS-Co$^{3+}$
splits into two singlets (L=0), and the triplet $t_{2g}$ into a singlet and a
doublet (L=1). The singlet states of the $t_{2g}$ level are occupied by two
electrons with opposite spins and the doublet level by three electrons. Due to
the degeneracy of the doublet level with nonzero angular momentum, a strong
intra-atomic spin-orbit coupling is created which leads to the increase in
magnetic anistropy, in contrast to the reference manganese perovskites. Our
result can also be interpreted as another evidence in favour of the orbital
instability of the Co ions under external magnetic fields in 
La$_{0.7}$Sr$_{0.3}$CoO$_3$.

\acknowledgements

Spanish MCYT, under project FEDER MAT2001-3749, is acknowledged. We also wish to
thank L. E. Hueso for data on La$_{0.7}$Ca$_{0.3}$MnO$_3$.

\begin{figure}  
\caption
{Magnetization vs. magnetic field in the saturation
regime of La$_{0.7}$Sr$_{0.3}$CoO$_3$. In fields up to 55 kOe the material is
still far from saturation, indicating a strong magnetic anisotropy.}       
\label{un}    
\end{figure}

\begin{figure}  
\caption
{(a) Representation of the magnetization vs. field
isotherms of La$_{0.7}$Sr$_{0.3}$CoO$_3$ according to Eq. 2. (b) Values
of $\chi$ used for the representations of (a). (c) Saturation
magnetization obtained from the fits of the isotherms to Eq. 1.}       
\label{dous}     
\end{figure}

\begin{figure}  
\caption{Magnetic anisotropy vs. temperature for
La$_{0.7}$Sr$_{0.3}$CoO$_3$, extracted from fits of the high field linear
part of magnetization vs. field curves (filled symbols) as well as from data of
coercive fields (open symbols).}        
\label{tres}   
\end{figure}

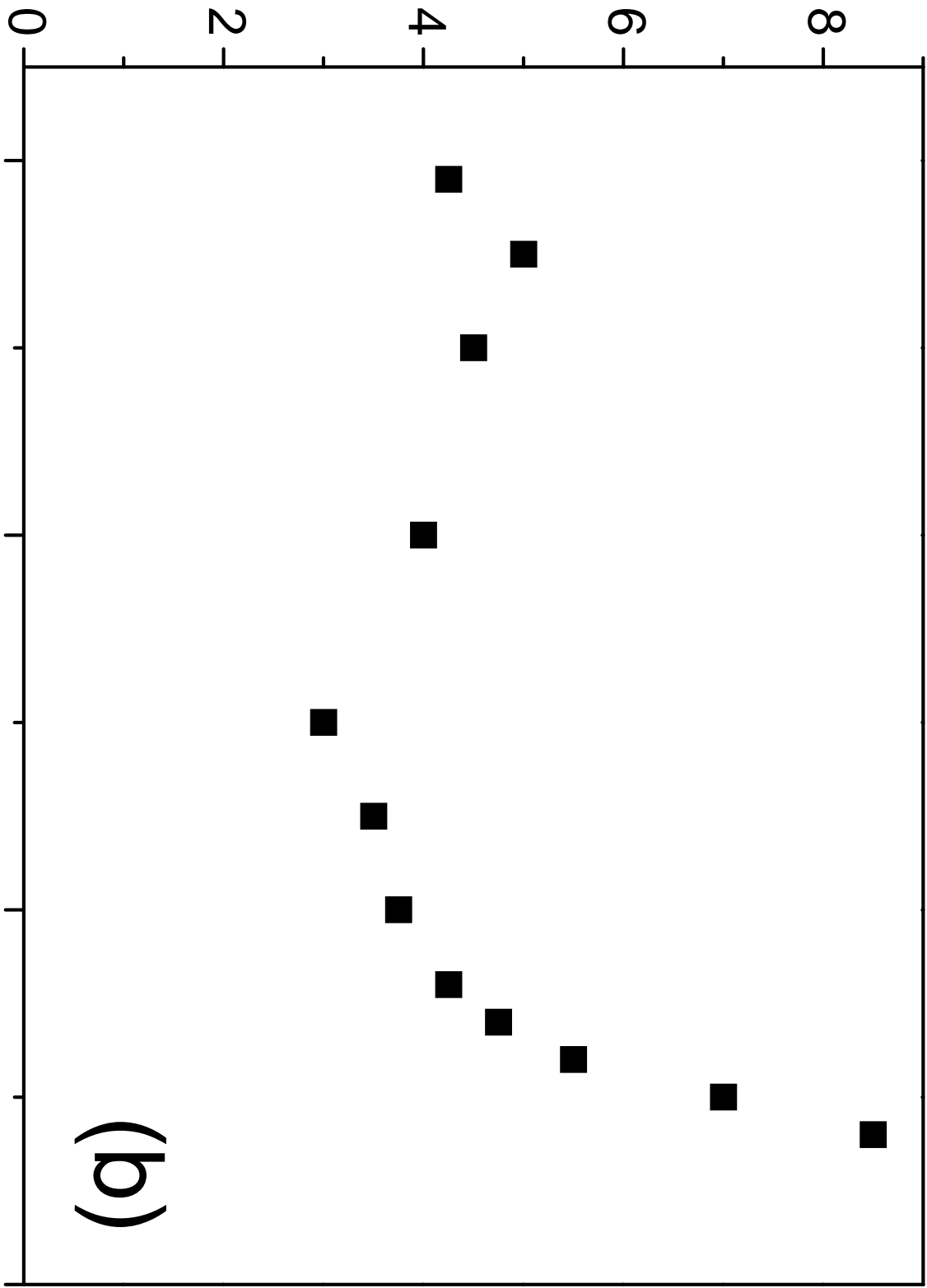
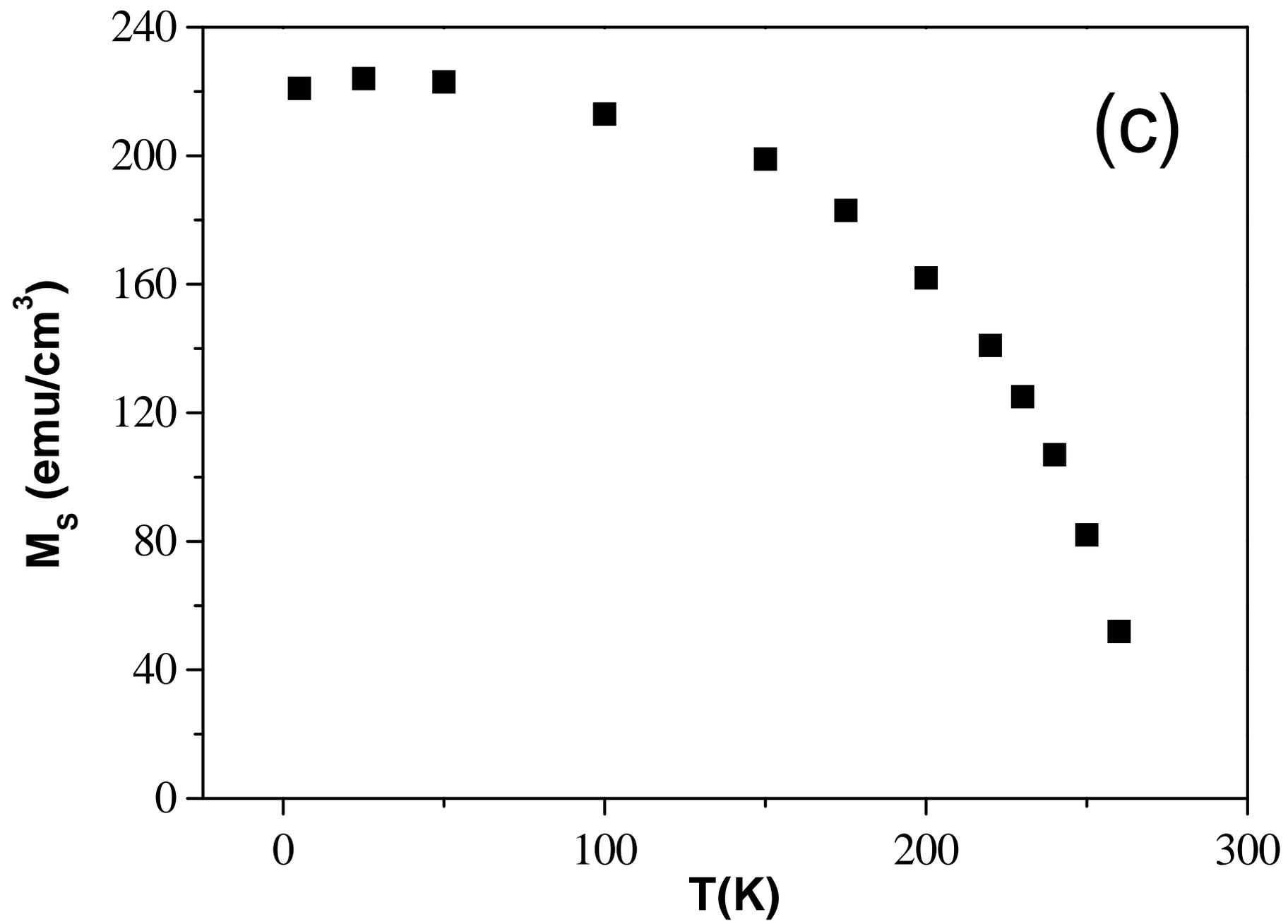
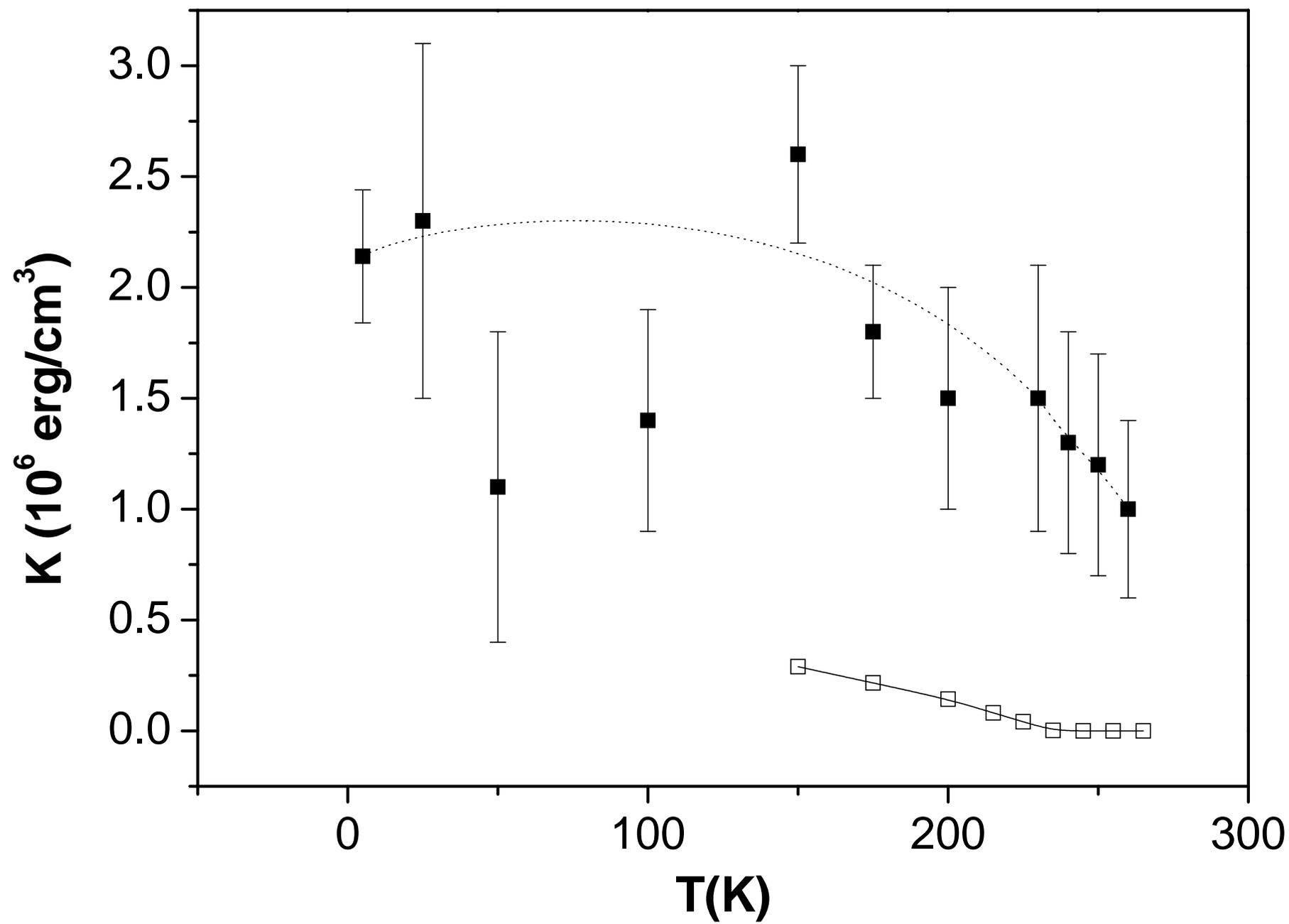
\end{document}